\documentclass[twocolumn,showpacs,preprintnumbers,amsmath,amssymb,prb]{revtex4}

\usepackage{graphicx}
\usepackage{dcolumn}
\usepackage{bm}
\usepackage{subfigure}
\usepackage{color}
\usepackage{amssymb}
\newcommand{\RAA}{\AA$^{-1}$}

\begin{document}

\preprint{APS/123-QED}

\title{Quantitative size-dependent structure and strain determination of CdSe nanoparticles
 using atomic pair distribution function analysis}

\author{A.~S. Masadeh, E.~S. Bo\v zin, C. L. Farrow, G. Paglia, P.~Juhas}

\author{S.~J.~L. Billinge}%
 \email{billinge@pa.msu.edu}

\affiliation{ Department of Physics and Astronomy, Michigan State
University, East Lansing, Michigan 48824-1116, USA }%

\author{A. Karkamkar }
\author{M.~G. Kanatzidis}

\affiliation{Department of Chemistry, Michigan State University,
East Lansing, Michigan 48824-1116, USA}


\begin{abstract}
The size-dependent structure of CdSe nanoparticles, with diameters
ranging from 2 to 4~nm, has been studied using the atomic pair
distribution function (PDF) method. The core structure of the
measured CdSe nanoparticles can be described in terms of the
wurtzite atomic structure with extensive stacking faults.  The
density of faults in the nanoparticles $\sim 50$\% . The diameter of
the core region was extracted directly from the PDF data and is in
good agreement with the diameter obtained from standard
characterization methods suggesting that there is little surface
amorphous region. A compressive strain was measured in the Cd-Se
bond length that increases with decreasing particle size being 0.5\%
with respect to bulk CdSe for the 2~nm diameter particles. This
study demonstrates the size-dependent quantitative structural
information that can be obtained even from very small nanoparticles
using the PDF approach.

\end{abstract}

\pacs{61.46.Df, 61.10.-i, 78.66.Hf, 61.46.-w }

\maketitle

\section{\label{sec:level1}Introduction}

Semiconductor nanoparticles are of increasing interest for both
applied and fundamental research. Wurtzite-structured cadmium
selenide is an important II-VI semiconducting compound for
optoelectronics.~\cite{hodes;prb87} CdSe quantum dots are the most
extensively studied quantum nanostructure because of their size-tunable
properties, and they have been used as a model system for
investigating a wide range of nanoscale electronic, optical,
optoelectronic, and chemical processes.~\cite{tolbe;s94} CdSe also
provided the first example of self-assembled semiconductor
nanocrystal superlattices.~\cite{murra;s95} With a direct band gap of
1.8~eV, CdSe quantum dots have been used for laser
diodes~~\cite{colvi;n94}, nanosensing~~\cite{tran;pss02} , and
biomedical imaging.~\cite{bruch;s98} In fundamental research,
particles with a diameter in the 1-5~nm range are of particular
importance since they cover the transition regime between the bulk
and molecular domains where quantum size effects play an important
role. Significant deviation from bulk properties are expected for
particles with diameter below 5~nm, and were observed in many
cases~~\cite{bruch;s98,li;prl03} as well as in this study.

Accurate determination of atomic scale structure, homogeneous and
inhomogeneous strain, structural defects and geometrical particle
parameters such as diameter and shape, are important for understanding
the fundamental mechanisms and processes in nanostructured
materials. However, difficulties are experienced when standard methods
are applied to small nanoparticles.  In this domain the presumption of
a periodic solid, which is the basis of a crystallographic analysis,
breaks down.  Quantitative determinations of the nanoparticle
structure require methods that go beyond crystallography. This was
noted early on in a seminal study by Bawendi {\it et
al.}~\cite{bawen;jcp89} where they used the Debye equation, which is
not based on a crystallographic assumption, to simulate
semi-quantitatively the scattering from some CdSe nanoparticles.
However, despite the importance of knowing the nanoparticle structure
quantitatively with high accuracy, this work has not been followed up
with application of modern local structural
methods~\cite{egami;b;utbp03,billi;cc04} until
recently.~\cite{gilbe;s04,neder;jpcm05,howel;prb06,page;chpl04,petko;jmc05,gates;prb05}
In this study we return to the archetypal CdSe nanoparticles to
investigate the extent of information about size-dependent structure
of nanoparticles from the atomic pair distribution function (PDF)
method.  This is a local structural technique that yields quantitative
structural information on the nanoscale from x-ray and neutron powder
diffraction data.~\cite{billi;cc04} Recent developments in both data
collection~\cite{chupa;jac03,proff;apa01i} and
modeling~\cite{proff;jac99,tucke;jpcm01} make this a potentially
powerful tool in the study of nanoparticles.  Additional extensions to
the modelling are necessary for nanoparticles, and some of these have
been successfully
demonstrated.~\cite{gilbe;s04,neder;jpcm05,kumpf;jcp05}

In this paper, we present a detailed analysis of the structural
information available from PDF data on (2-4~nm) CdSe
nanoparticles. The PDF method is demonstrated here as a key tool that
can yield precise structural information about the nanoparticles such
as the atomic structure size of the core, the degree of crystallinity,
local bonding, the degree of the internal disorder and the atomic
structure of the core region, as a function of the nanoparticle
diameter. Three CdSe nanoparticle samples with different diameters
that exhibit different optical spectra have been studied. The purpose
of this paper is not only to explain the PDF data of CdSe
nanoparticles through a modeling process, but also to systematically
investigate the sensitivity of the PDF data to subtle structural
modifications in nanoparticles relative to bulk material.

The measurement of the nanoparticle size can lead to significantly
different results when performed by different methods, and there is no
consensus as to which is the most
reliable.~\cite{wickh;prl00,nanda;cm00} It is also not clear that a
single diameter is sufficient to fully specify even a spherical
particle since the presence of distinct crystalline core and
disordered surface regions have been postulated.~\cite{bawen;jcp89}

Powder diffraction is a well established method for structural and
analytical studies of crystalline materials, but the applicability to
such small particles of standard powder diffraction based on
crystallographic methods is questionable and likely to be
semi-quantitative at best. Palosz~{\it et al.}~\cite{palos;appa02} have
shown that the conventional tools developed for elaboration of powder
diffraction data are not directly applicable to
nanocrystals.~\cite{palos;appa02} There have been some
reports~\cite{borch;l05,nanda;cm00,bawen;jcp89} in the past few years
extracting nanoparticle diameter from x-ray diffraction (XRD) using
the Scherrer formula, which is a phenomenological approach that
considers the finite size broadening of
Bragg-peaks.~\cite{guini;b;xdic63} This approach will decrease in
accuracy with decreasing particle size, and for particle sizes in the
range of a few nanometers the notion of a Bragg peak becomes moot.~\cite{palos;appa02}  At
this point the Debye formula~\cite{debye;ap15} becomes the more
appropriate way to calculate the scattering.~\cite{bawen;jcp89} The
inconsistency between the nanoparticle diameter determined from the
standard characterization methods and the diameter obtained by
applying the Scherrer formula have been observed by several
authors.~\cite{bawen;jcp89,nanda;cm00,kaszk;zk06}

Previous studies of CdSe nanoparticle structure have demonstrated the
sensitivity of the XRD pattern to the presence of planar disorder and
thermal effects due to nano-size
effects.~\cite{murra;jacs93,bawen;jcp89} The diffraction patterns of
CdSe nanoparticles smaller than 2.0~nm have been observed to appear
markedly different from those of the larger diameters (see
Ref.~\onlinecite{murra;jacs93} Fig.~11), the large attenuation and
broadening in the Bragg reflections in these small nanoparticles,
making the distinction between wurtzite and zinc-blende hard using
conventional XRD methods. Murray~{\it et al.}~\cite{murra;jacs93}
reported that the combination of X-ray studies and TEM imaging yields
a description of the average CdSe nanoparticle structure. Strict
classification of the CdSe nanoparticles structure as purely wurtzite
or zinc-blend is potentially misleading.~\cite{murra;jacs93}
Bawendi~{\it et al.}~\cite{bawen;jcp89} reported that CdSe
nanoparticles are best fit by a mixture of crystalline structures
intermediate between zinc-blend and wurtzite.  Here we apply the PDF
method to CdSe nanoparticles and refine quantitative structural
parameters to a series of CdSe nanoparticles of different sizes.

Strain in nano systems has been observed before in different studies,
as well as in this study.  Using combined PDF and extended
X-ray-absorbtion fine structure (EXAFS) methods, Gilbert{\it et
al.}~\cite{gilbe;s04} observed a compressive strain compared to the
bulk in ZnS nanocrystals.  Using an electric field-induced resonance
method, Chen~{\it et al.}~\cite{chen;prl06} detected the enhancement
of Young's modulus of ZnO nanowires along the axial direction when the
diameters are decreased. Very recently, Quyang~{\it et
al.}~\cite{ouyan;apl06} developed an analytical model for the
size-induced strain and stiffness of a nanocrystal from the
perspective of thermodynamics and a continuum medium approach. It was
found theoretically that the elastic modulus increases with the
inverse of crystal size and vibration frequency is higher than that of
the bulk.~\cite{ouyan;apl06} Experimentally, the Cd$Q$~$(Q$=$S,Se,Te)$
first-neighbor distances have been studied using both XRD and EXAFS
methods.~\cite{marcu;nsm92} The distances were found smaller than
those in the bulk compounds by less than 1.0\%.  Herron~{\it et
al.}~\cite{herro;s93} studied CdS nanocrystals and showed a bond
contraction of $\sim 0.5$\% compared to the bulk. Carter~{\it et
al.}~\cite{carte;prb97} studied a series of CdSe nanoparticles using
the EXAFS method. In the first shell around both the Se and Cd atoms,
they found essentially no change in the first-neighbor
distance. Chaure~{\it et al.}~\cite{chaur;pe05} studied the strain in
nanocrystalline CdSe thin films, using Raman scattering and observed a
peak shift with decrease in particle size, which was attributed to the
increase in stress with decreasing particle size.~\cite{chaur;pe05}

 Local structural deviations or disorder mainly affect the diffuse
scattering background.  The XRD experiments probe for the presence of
periodic structure which are reflected in the Bragg peaks. In order to
have information about both long-range order and local structure
disorder, a technique that takes both Bragg and diffuse scattering
need to be used, such as the PDF technique.  Here we apply the PDF
method to study the structure, size and strain in CdSe nanoparticles
as a function of nanoparticle diameter.  The core structure of the
CdSe nanoparticles can be described by a mixture of crystalline
structures intermediate between zinc-blend and wurtzite, which is
wurtzite containing a stacking fault density (SFD) of up to $\sim
50$\%, with no clear evidence of a disordered surface region,
certainly down to 3~nm diameter.  The structural parameters are
reported quantitatively.  We measure a size-dependent strain on the
Cd-Se bond which reaches 0.5\% at the smallest particle size.  The
size of the well-ordered core extracted directly from the data agrees
with the size determined from other methods.

\section{experimental details }
\subsection{Sample preparation}
CdSe nanoparticles were synthesized from cadmium acetate, selenium,
trioctyl phosphine and trioctyl phosphine oxide.  Sixty four grams of
trioctylphosphine oxide (TOPO) containing cadmiumacetate was heated to
360$^\circ$C under flowing argon. Cold stock solution (38.4~ml) of
(Se:trioctylphosphine~=~2:100 by mass) was quickly injected into the
rapidly stirred, hot TOPO solution. The temperature was lowered to
300$^\circ$C by the injection. At various time intervals, 5-10~ml
aliquots of the reaction mixture were removed and precipitated in
10~ml of methanol. The color of the sample changed from bright yellow
to orange to red to brown with time interval variation from 20 seconds
to 1200 seconds.  Three nanoparticle sizes, CdSeI (small), CdSeII
(medium) and CdSeIII (large), were used for this study, as well as a
bulk CdSe sample for reference.

\begin{figure}[tbp]
\includegraphics[ width=0.45\textwidth]{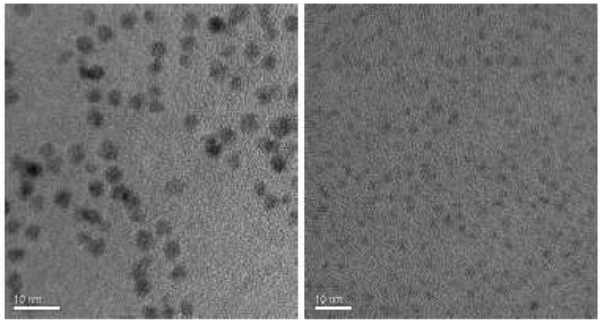}
\caption{TEM image of CdSe nanocrystal prepared using the method
described in the text. CdSe obtained by 1200~seconds (left) and 15~seconds
(right) nucleation. The line-bar is 10~nm in size in both images.
} \label{fig:cdse1200tem}
\end{figure}

\begin{figure}[tbp]
\includegraphics[ width=0.45\textwidth]{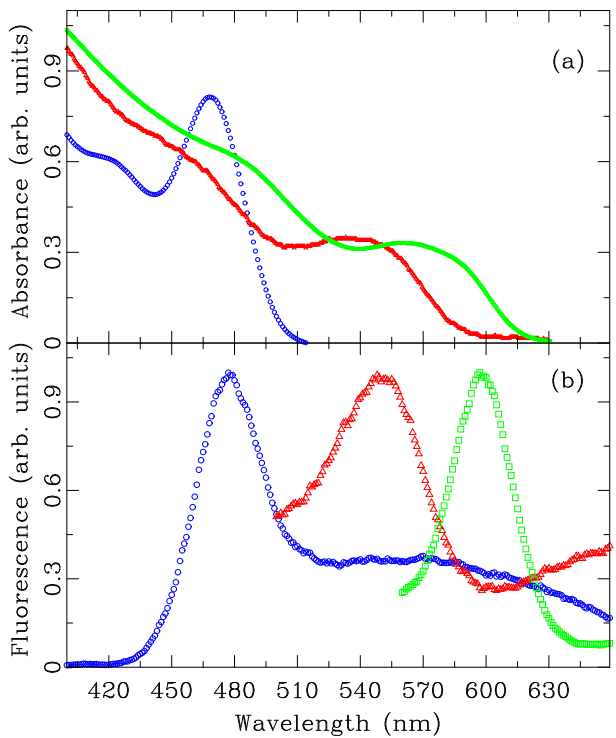}
\caption{(a) Room temperature UV-vis absorption and (b)
   photoluminescence spectra from the sample of CdSe
   nanocrystals. (\textcolor{blue}{$\bigcirc$}) CdSeI,
   (\textcolor{red}{$\triangle$}) CdSeII,
   (\textcolor{green}{$\square$}) CdSeIII.} \label{fig:uv_abs}
\end{figure}

The samples were further purified by dissolving and centrifuging in
methanol to remove excess TOPO. This process also resulted in a
narrower particle size distribution. The transmission electron
micrograph (TEM) images (Fig.~\ref{fig:cdse1200tem}) show uniformly
sized
nanoparticles with no signs of aggregation.  The ultraviolet
visible (UV-vis) absorption and photoluminescence (PL) spectra of
the aliquots were recorded by redissolving the nanocrystals in
toluene. The spectra are shown in Fig.~\ref{fig:uv_abs}.

 The band-gap values obtained for the measured samples can be
correlated with the diameter of the nanoparticles based on the table
provided in supplementary information of Peng~{\it et al.}~\cite{peng;jacs98}
using the data on exciton peaks measured with UV-visible light
absorption, and photoluminescence peaks. The particle sizes were
measured by TEM as well. The measured values of particle diameter
using these various methods are summarized  in
Table~\ref{tab;NPZtemPDF}.
\begin{table}[tb]
\caption{ CdSe nanoparticle diameter as determined using various methods. }
\begin{tabular}{lcccccccc}  
\hline \hline
   ~~  &  CdSeIII  &~~~ CdSeII ~~~ & CdSeI   \\
 \hline
 Nucleation time (s)  ~~   & 1200  &  630   &  15  \\
 Diameter (nm)  ~~   &     &      &           \\
 TEM     & ~3.5(2)   & ~2.7(2) &  ~2.0(2)    \\
 UV-vis  & ~3.5(4)   & ~2.9(3) &  ~$\le$ 1.90 \\
 PL      & ~3.6(4)   & ~2.9(3) &  ~$\le$ 2.1   \\
 PDF     & ~3.7(1)   & ~3.1(1) &  ~2.2(2)       \\
\hline \hline
\end{tabular}
\label{tab;NPZtemPDF}
\end{table}
%

\subsection{The atomic PDF method}
The atomic PDF analysis of x-ray and neutron powder diffraction
data is a powerful method for studying the structure of
nanostructured
materials.~\cite{billi;cc04,egami;b;utbp03,petko;jacs00,petko;jacs02,petko;prb04,juhas;n06}
Recently, it has been explicitly applied to study the structure of
discrete
nanoparticles.~\cite{zhang;n03,gilbe;s04,neder;jpcm05,petko;prb05,juhas;n06}
The PDF method can yield precise structural and size information,
provided that special care is applied to the measurement and to
the method used for analyzing the data. The atomic PDF, $G(r)$, is
defined as
\begin{equation}
\centering \label{equ;eqGrf} G\left(r\right) = 4\pi
r\left[\rho\left(r\right)-\rho_{0}\right],
\end{equation}
where $\rho(r)$ is the atomic pair-density, $\rho_{0}$ is the
average atomic number density and~$r$ is the radial
distance.~\cite{warre;b;xd90}  The PDF yields the probability of
finding pairs of atoms separated by a distance~$r$. It is obtained
by a sine Fourier transformation of the reciprocal space total
scattering structure function $S(Q)$, according to
\begin{equation}
\centering \label{equ;sqtogr} G\left(r\right) =
{2\over\pi}\int_0^{\infty}Q[S(Q)-1]\sin Qr\>\mathrm{d}Q,
\end{equation}
where $S(Q)$ is obtained from a diffraction experiment.
 This approach is widely used for studying liquids,
amorphous and crystalline materials, but has recently also been
successfully applied to nanocrystalline
materials.~\cite{billi;cc04}


\subsection{High-energy x-ray diffraction experiments}

X-ray powder diffraction experiments to obtain the PDF were performed
at the 6IDD beamline at the Advanced Photon Source at Argonne National
Laboratory. Diffraction data were collected using the recently
developed rapid acquisition pair distribution function (RAPDF)
technique~\cite{chupa;jac03} that benefits from 2D data
collection. Unlike TEM, XRD probes a large number of crystallites that
are randomly oriented.  The powder samples were packed in a flat plate
with thickness of 1.0~mm sealed between kapton tapes. Data were
collected at room temperature with an x-ray energy of 87.005~keV
($\lambda = 0.14248$~\AA).  An image plate camera (Mar345) with a
diameter of 345~mm was mounted orthogonally to the beam path with a
sample to detector distance of 208.857~mm, as calibrated by using
silicon standard sample.~\cite{chupa;jac03} The image plate was exposed
for 10~seconds and this was repeated 5 times for a total data
collection time of 50~seconds.  The RAPDF approach avoids detector
saturation whilst allowing sufficient statistics to be obtained. This
approach also avoids sample degradation in the beam that was observed
for the TOPO coated nanoparticles during longer exposures, on the
scale of hours, that were required using conventional point-detector
approaches. To reduce the background scattering, lead shielding was
placed before the sample with a small opening for the incident beam.

\begin{figure}[tbp]
\includegraphics[width=0.22\textwidth]{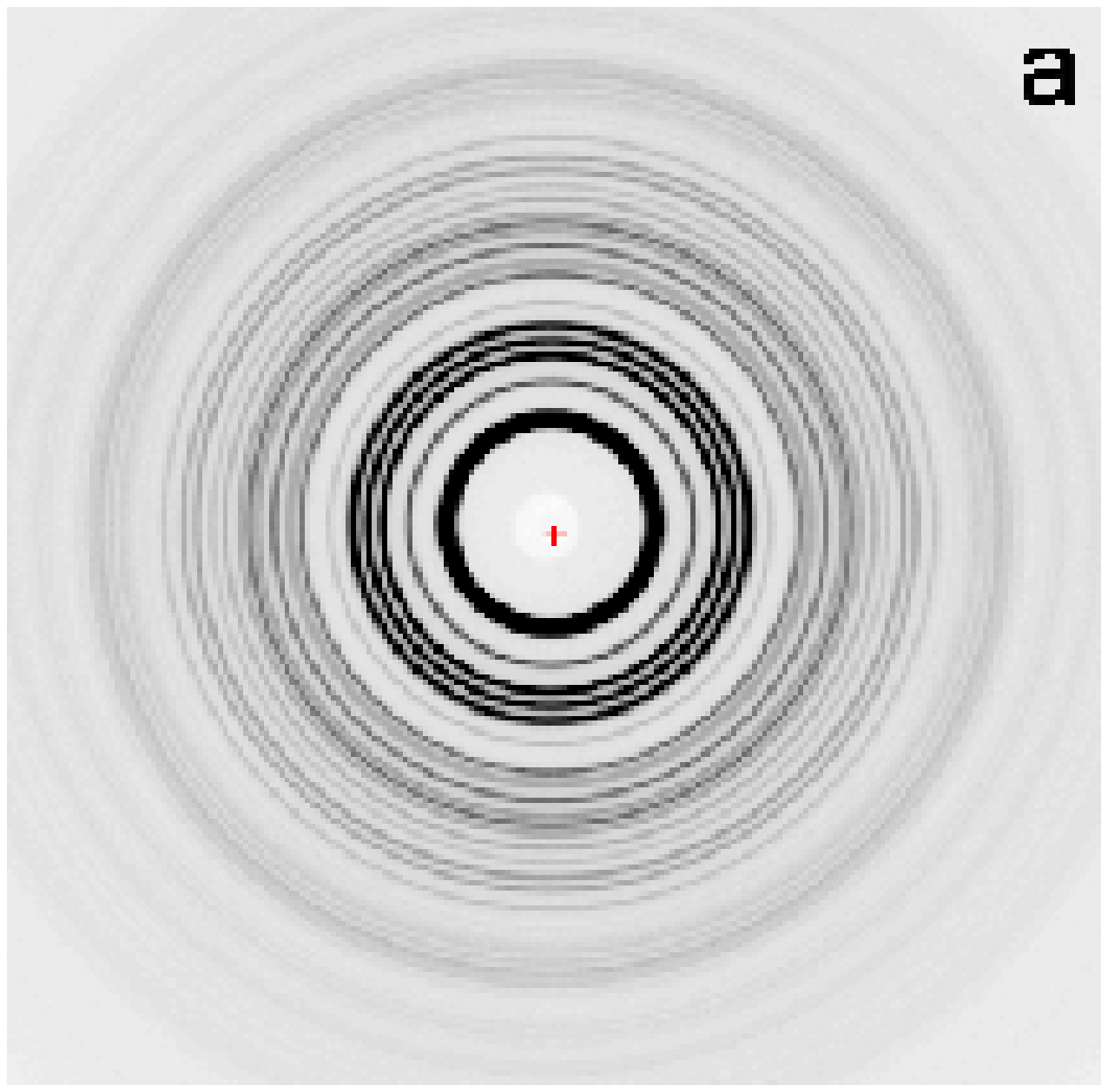}
\includegraphics[width=0.22\textwidth]{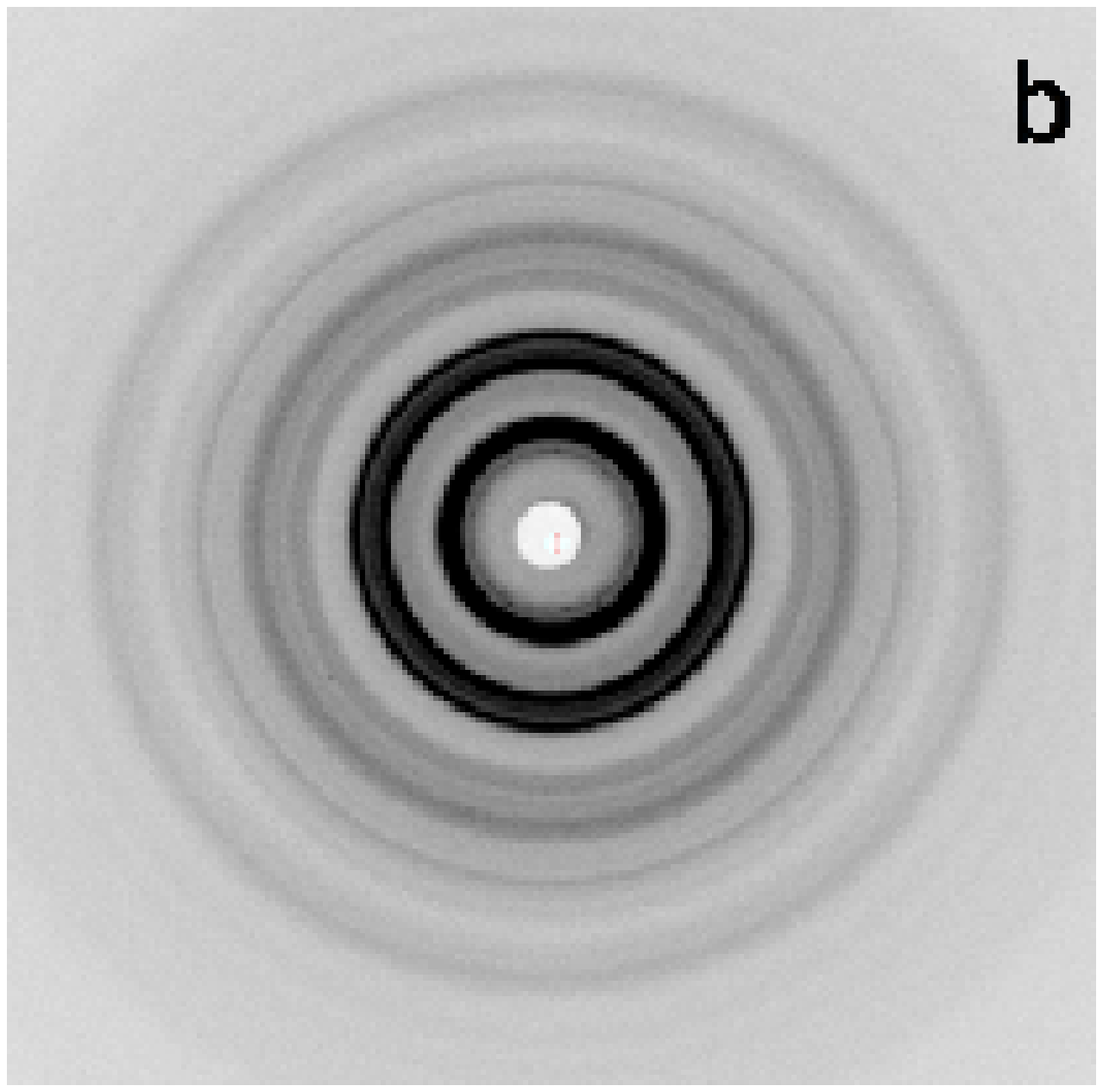}
\caption {Two dimensional XRD raw data collected using image plate
detector from (a) CdSe bulk and (b) nanoparticle  CdSeIII samples. }
\label{fig:2drawdata1}
\end{figure}

\begin{figure}[tbp]
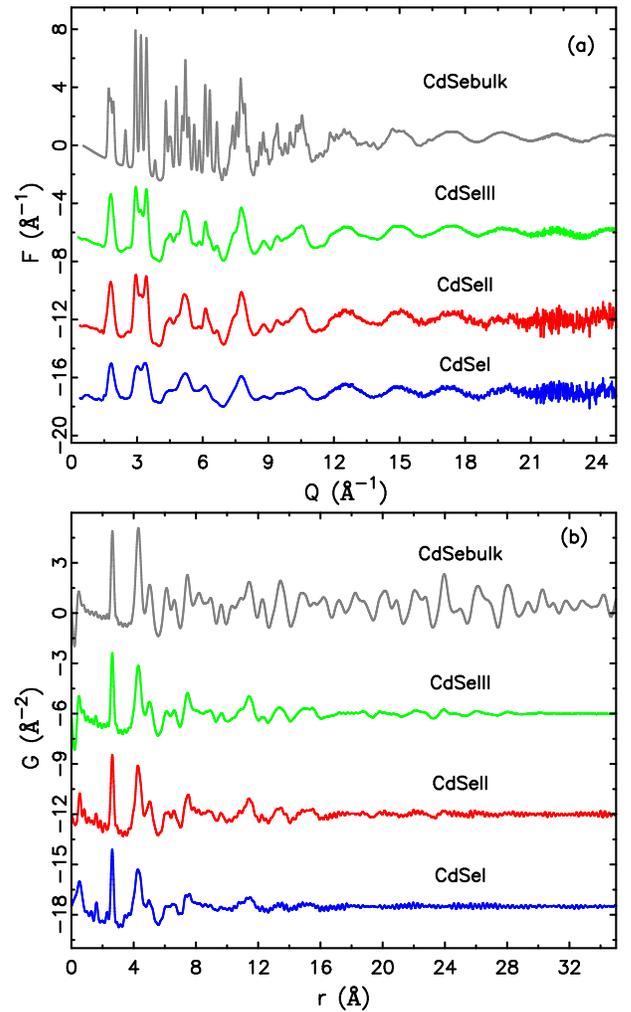

\includegraphics[width=0.45\textwidth]{fig4_a.ps}
\includegraphics[width=0.45\textwidth]{fig4_b.ps}
 \caption {(a) The experimental reduced structure function $F(Q)$ of
CdSe nanoparticle with different diameters and (b) the corresponding
PDF, $G(r)$, obtained by Fourier transformation of the data in (a)
with $Q_{max}$ = 25.0~\RAA, from top to bottom: bulk, CdSeIII,
CdSeII and CdSeI.  } \label{fig:allrawfqgr}
\end{figure}

Examples of the raw 2D data are shown in Fig.~\ref{fig:2drawdata1}.
These data were integrated and converted to intensity versus
2$\theta$ using the software Fit2D,~\cite{hamme;hpr96} where
2$\theta$ is the angle between the incident and scattered x-ray
beam.  The integrated data were normalized by the average monitor
counts. The data were corrected and normalized~\cite{egami;b;utbp03}
using the program PDFgetX2~\cite{qiu;jac04i} to obtain the total
scattering structure function, $S(Q)$, and the PDF, $G(r)$, which
are shown in Figs.~\ref{fig:allrawfqgr} (a) and (b) respectively.
The scattering signal from the surfactant (TOPO) was measured
independently and subtracted as a background in the data reduction.

In the Fourier transform step to get from $S(Q)$ to the PDF $G(r)$,
the data are truncated at a finite maximum value of the momentum
transfer, $Q=Q_{max}$.  Different values of Q$_{max}$ may be chosen.
Here a Q$_{max}= 25.0$~\RAA\  was found to be optimal. Q$_{max}$ is
optimized such as to avoid large termination effects and to
reasonably minimize the introduced noise level as signal to noise
ratio decreases with $Q$ value.

Structural information was extracted from the PDFs using a
full-profile real-space local-structure refinement
method~\cite{billi;b;lsfd98} analogous to Rietveld
refinement.~\cite{rietv;jac69} We used an updated
version~\cite{farro;jpcm07} of the program PDFfit~\cite{proff;jac99} to
fit the experimental PDFs.  Starting from a given structure model and
given a set of parameters to be refined, PDFfit searches for the best
structure that is consistent with the experimental PDF data.  The
residual function ($R_w$) is used to quantify the agreement of the
calculated PDF from model to experimental data:
\begin{equation}
R_w = \sqrt {\frac{\sum_{i=1}^{N}\omega
(r_{i})[G_{obs}(r_{i})-G_{calc}(r_{i})]^2}{\sum_{i=1}^{N}\omega
(r_{i})G_{obs}^{2}(r_{i})}}.
\end{equation}
Here the weight $\omega (r_{i})$ is set to unity which is justified
because in $G(r)$ the statistical uncertainty on each point is
approximately equal.~\cite{toby;aca04,toby;aca92}

 The structural parameters of the model were unit cell parameters,
anisotropic atomic displacement parameters (ADPs) and the fractional
coordinate~{\it z}~ of Se/Cd atom. Non structural parameters that were
refined were a correction for the finite instrumental resolution,
($\sigma_{Q}$), low-$r$ correlated motion peak sharpening factor
($\delta$),~\cite{jeong;jpc99,jeong;prb03} and scale factor.  When
estimating the particle size, a new version of the fitting program
with particle size effects included as a refinable
parameter~\cite{farro;unpub06} was used. The sample resolution
broadening was determined from a refinement to the crystalline CdSe
and the silicon standard sample and fixed and the particle diameter
refined, as described below. Good agreement between these results was
obtained.

\section{results and discussion}

The reduced structure functions for the bulk and nanocrystalline
samples are shown plotted over a wide range of $Q$ in
Fig~\ref{fig:allrawfqgr}(a).  All of the patterns show significant
intensity up to the highest values of $Q$, highlighting the value
of measured data over such a wide $Q$-range.  All of the
diffraction patterns have peaks in similar positions reflecting
the similarity of the basic structures, but as the nanoparticles
get smaller the diffraction features become broadened out due to
finite size effects.~\cite{guini;b;xdic63}

The PDFs are shown in Fig.~\ref{fig:allrawfqgr}(b).  What is apparent
is that, in real-space, the PDF features at low-$r$ are comparably
sharp in all the samples.  The finite size effects do not broaden
features in real-space.  The finite particle size is evident in a
fall-off in the intensity of structural features with increasing-$r$.
Later we will use this to extract the average particle size in the
material.  The structure apparent in the $G(r)$ function comes from
the atomic order within the nanoparticle.  The value of~$r$ where
these ripples disappear indicates the particle core region diameter;
or at least the diameter of any coherent structural core of the
nanoparticle. By direct observation (Fig.~\ref{fig:NPZ}) we can put a
lower limit on the particle diameters to be 3.6, 2.8 and 1.6~nm for
CdSeIII, II and I, respectively, where the ripples can be seen to die
out by visual inspection. These numbers will be quantified more
accurately later.
\subsection{Nanoparticle structure}
 Features in the PDF at low-$r$ reflect the internal structure of the
nanoparticles.  The nanoparticle PDFs have almost the same features as
in the bulk in the region below 8.0~\AA , reflecting the fact that
they share a similar atomic structure on average. In the finite
nano-size regime, local structural deviations from the average bulk
structure are expected. 

 A large number of semiconductor alloys, especially some sulfides and
selenides, do not crystallize in the cubic zinc-blende structure but
in the hexagonal wurtzite structure~\cite{kikum;jcg85}. Both wurtzite
and zinc-blende structures are based on the stacking of identical
two-dimensional planar units translated with respect to each other, in
which each atom is tetrahedrally coordinated with four nearest
neighbors. The layer stacking is described as {\it ABABAB...} along
the [001] axis for wurtzite and as {\it ABCABC...}  along the [111]
axis for zinc-blende.
\begin{figure}[tbp]
\includegraphics[width=0.45\textwidth]{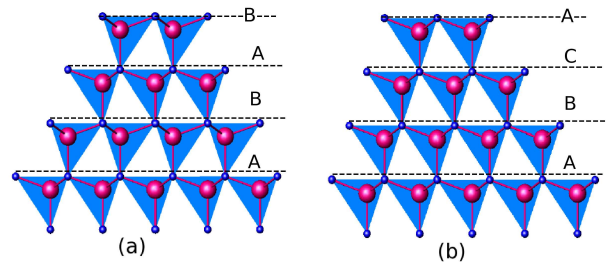}
\caption {Fragments from the (a) wurtzite structure, space group
($P6_{3}mc$) and (b) zinc-blende structure, space group
($F\bar43m$). } \label{fig:Wur-ZB-fragments}
\end{figure}
As can be seen in the Fig.~\ref{fig:Wur-ZB-fragments}, each
cadmium and selenium is tetrahedrally coordinated in both
structures. However, the next nearest and more distant coordination
sequences are different in the two structures.

The largest changes in structure are expected in the smallest
nanoparticles. In these small nanoparticles, the proportion of atoms
on the surface is large making the notion of a well-ordered
crystalline core moot. The fraction of atoms involved in the surface
atoms was estimated as 0.6, 0.45 and 0.35 for 2~nm, 3~nm and 4~nm
nanoparticle diameters, respectively. This was estimated by taking
different spherical cuts from bulk structure, then counting the atom
with coordination number 4 as core atom and the one with less than 4
as surface atom.  For the smallest particles the small number of atoms
in the core makes it difficult to define a core crystal structure,
making the distinction between wurtzite and zinc-blende difficult
using the conventional XRD methods as nanoparticle size
decreases.~\cite{murra;jacs93} The principle difference between these
structures is the topology of the CdSe$_4$ connections, which may also
be becoming defective in the small nanoparticles.

\begin{figure}[tbp]
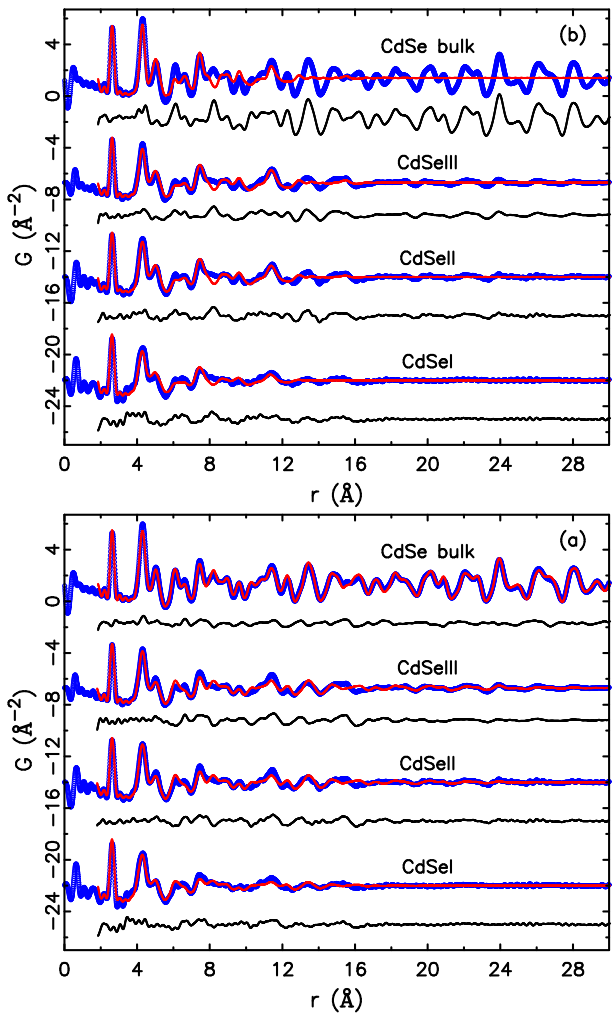

\includegraphics[width=0.45\textwidth]{fig6_a.ps}
\hfill
\includegraphics[width=0.45\textwidth]{fig6_b.ps}
\caption {(Color online) The experimental PDF, $G(r)$, with $Q_{max}$ = 19.0~\RAA
 (blue solid dots) and the calculated PDF from refined structural
 model (red solid line), with the difference curve offset below (black
 solid line). PDF data are fitted using (a) wurtzite structure model,
 space group $P6_{3}mc$ and (b) zinc-blende model with space group
 $F\bar43m$. In both models isotropic atomic displacement factors  (U$_{iso}$) are used. }
\label{fig:coreFITiso}
\end{figure}

Two structure models wurtzite (space group $P6_{3}mc$) and zinc-blende
(space group $F\bar43m$), were fit to the PDF data.  The results of
the full-profile fitting to the PDF data are shown
Fig.~\ref{fig:coreFITiso}. In this figure we compare fits to the (a)
wurtzite and (b) zinc-blende structure models using isotropic atomic
displacement factors (U$_{iso}$) in both models. The wurtzite
structure gives superior fits for the bulk structure. However, for all
the nanoparticle sizes, the fits of wurtzite and zinc-blende are
comparable as evident from the difference curves in
Fig.~\ref{fig:coreFITiso} and the R$_{w}$-values reported in
Table~\ref{tab;rwpiso}.  This indicates that classification of the
CdSe nanoparticles structure as purely wurtzite or zinc-blend is
misleading~\cite{murra;jacs93} and it is better described as being
intermediate between the two structures, as has been reported
earlier~\cite{bawen;jcp89}.

\begin{figure}[tbp]
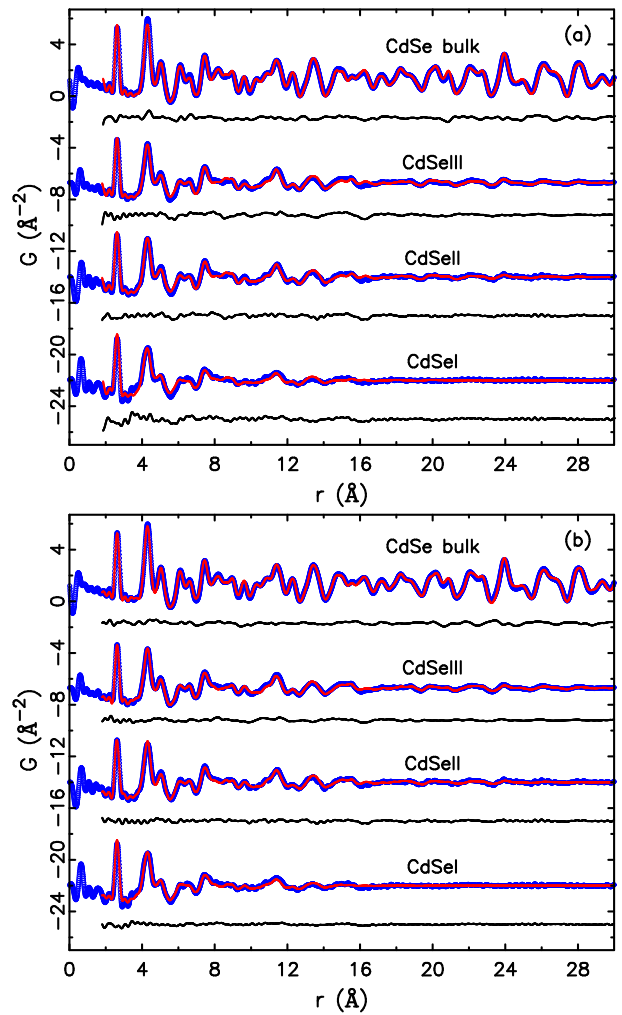

\includegraphics[width=0.45\textwidth]{fig7_a.ps}
\hfill
\includegraphics[width=0.45\textwidth]{fig7_b.ps}
\caption {(Color online) The experimental PDF, $G(r)$, with $Q_{max}$
 = 19.0~\RAA (blue solid dots) and the calculated PDF from refined
 structural model (red solid line), with the difference curve offset
 below (black solid line). PDF data are fitted using wurtzite
 structure model (a) with no stacking fault and (b) with 33\% stacking
 fault density for bulk and 50\% for all nanoparticle sizes. In both
 cases anisotropic atomic displacement factors  (U$_{aniso}$) are used } \label{fig:coreFITsfd}
\end{figure}
\begin{table}
\caption{The refined residual (R$_{w}$) values obtained from PDF
 analysis assuming the wurtzite and zinc-blend structure models with
 space group $P6_{3}mc$ and $F\bar43m$, respectively.  In both models
 isotropic atomic displacement factors (U$_{iso}$) are used. }
\begin{tabular}{lcccccccc}  
\hline
 \hline
    & CdSe-bulk & CdSeIII &CdSeII & CdSeI   \\
 \hline
  Wurtzite (R$_{w}$)  &  0.16  & 0.31   &   0.28  &  0.31  \\
 \hline
  Zinc-blende (R$_{w}$)  &  0.52 &  0.32  &   0.30 &  0.35  \\
\hline

 \hline
\end{tabular}
\label{tab;rwpiso}
\end{table}
Introducing anisotropic ADPs (U$_{11}$ = U$_{22}$ $\neq $ U$_{33}$)
into the wurtzite model, resulted in better fits to the data.  The
refined parameters are reproduced in Table~\ref{tab;rwPDFwur} and the
fits are shown in Fig.~\ref{fig:coreFITsfd}(a). The values for the
nanoparticles are rather close to the values in the bulk wurtzite
structure. The model with anisotropic ADPs resulted in lower R$_{w}$.
There is a general increase in the ADPs with decreasing particle
size. This reflects inhomogeneous strain accommodation in the
nanoparticles as we discuss below. However, the values of the ADPs
along the $z$-direction for Se atoms (U$_{33}$) are four times larger
in the nanoparticles compared with the bulk where they are already
unphysically large.  The fact that this parameter is large on the Se
site and small on the Cd site is not significant, since we can change
the origin of the unit cell to place a Cd ion at the (1/3,2/3,z)
position and the enlarged U$_{33}$ shifts to the Cd site in this case.

The unphysically large U$_{33}$ value on the Se site is likely to be
due to the presence of faults in the basal plane stacking. For
example, similar unphysical enlargements of perpendicular thermal
factors in PDF measurements are explained by the presence of
turbostratic disorder in layered carbons\cite{petko;pmb99}, which is a
similar effect to faults in the {\it ABABAB} wurtzite stacking. Also,
the presence of stacking faults in the nanoparticles has been noted
previously.~\cite{bawen;jcp89} It is noteworthy that this parameter is
enlarged in EXAFS analyses of bulk wurtzite structures, probably for
the same reason.~\cite{marcu;nsm92,marcu;jpc91,rocke;pccp98} We suspect that the
enlargement in this parameter (U$_{33}$) is related to the stacking
fault density present in bulk and that is increasing in the
nanoparticles.

To test this idea we simulated PDF data using the wurtzite structure
containing different stacking fault densities. The stacking faults
were simulated for different  densities (0.167, 0.25,
0.333, and 0.5) by creating wurtzite superlattices with different
stacking sequences along the $C$-axis. The program DISCUS
\cite{proff;jac97} was used to create the stacking fault models and
PDFgui \cite{farro;jpcm07} was used to generate the corresponding
PDFs. The PDFs were simulated with all the ADPs fixed at
U$_{ii}=0.0133$~\AA$^{2}$, the value observed in the experimental bulk
data collected at room temperature (see Table~\ref{tab;rwPDFwur}).

\begin{figure}[tbp]
\includegraphics[ width=0.42\textwidth]{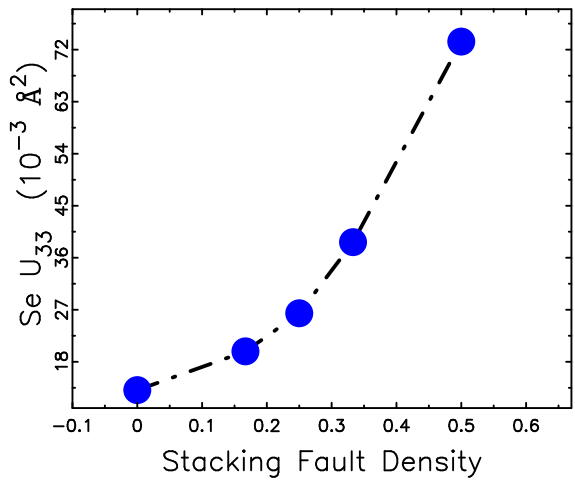}
\caption{The the enlargement in the the ADPs along the $z$-direction
for Se site U$_{33}$, as a function of the stacking fault density. }
\label{fig:sfd}
\end{figure}

To see if this results in enlarged U$_{33}$ values we refined the
simulated data containing stacking faults using the wurtzite model
without any stacking faults.  Indeed, the refined Se site U$_{33}$
increased monotonically with increasing stacking fault density. The
results are plotted in Fig.~\ref{fig:sfd}.

Fig.~\ref{fig:sfd} can be considered as calibration curve of
stacking fault density in the wurtzite structure, based on the
enlargement in the ADPs along the $z$-direction U$_{33}$.  From
this we can estimate a stacking fault density  of $\sim 35$\% for
our bulk CdSe sample, and $\sim 50$\% for each of the nanoparticles.

It is then possible to carry out a refinement using a structural 
model that contains an appropriate stacking fault density.  The PDF
data of bulk CdSe was therefore fit with a wurtzite model with a
33\% density, and the nanoparticle PDF fit with a model with 50\% of
stacking faults. The refinements give excellent fits, as is evident
in Fig.~\ref{fig:coreFITsfd}(b). The results are presented in
Table~\ref{tab;rwPDFwur}. The enlarged U$_{33}$ parameter on the Se
site is no longer present and it is now possible to refine
physically reasonable values for that parameter.  As well as
resulting in physically reasonable ADPs, the quality of the fits to
the data are excellent, though the $R_w$ value is slightly larger in
the nanoparticles.
\begin{table*}[tbp]
\caption{The refined parameters values obtained from PDF analysis
 assuming the wurtzite structure , space group $P6_{3}mc$, with
 different stacking fault densities (SFDs).  }
\begin{tabular}{lccccccccccccc}  
\hline \hline
\multicolumn{9}{c}{  ~~~~~~~~~~~~~~~~~~~~~~~~CdSe-bulk        ~~~~~~~~~~~~CdSeIII             ~~~~~~~~~~~~~~~~~~~ CdSeII     ~~~~~~~~~~~~~~~~CdSeI  } \\
\hline
Stacking fault density (\%)           & 0.0         &   33.0          & 0.0        & 50.0            & 0.0        &  50.0        & 0.0          &  50.0    \\
 a (\AA )                           & 4.3014(4)    & 4.3012(4)~     & 4.2997(9)  & 4.2987(9)~      & 4.3028(9)  & 4.3015(9) ~  & 4.2930(9)    & 4.2930(8)  \\
 c (\AA )                           & 7.0146(9)    & 7.0123(9)~     & 7.0145(4)  & 7.0123(4)~      & 6.9987(9)  & 6.9975(9)~   & 6.9405(9)    & 6.9405(7)  \\
 Se {\it Z-frac.}                    & 0.3774(3)    & 0.3771(3)~     & 0.3761(9)  & 0.3759(9)~      & 0.3751(6)  & 0.3747(6) ~  & 0.3685(9)    & 0.3694(9)    \\
 Cd U$_{11}$ = U$_{22}$ (\AA$^{2}$)  & 0.0108(2)    & 0.0102(2)~     & 0.0146(7)  & 0.0149(7) ~     & 0.0149(6)  & 0.0112(5)    & 0.0237(9)    & 0.0213(8)  \\
  ~~~~ U$_{33}$ (\AA$^{2}$)          & 0.0113(3)    & 0.0112(3) ~    & 0.0262(9)  & 0.0241(9)~      & 0.0274(9)  & 0.0271(9)~   & 0.0261(9)    & 0.0281(9)  \\
 Se ~U$_{11}$ = U$_{22}$ (\AA$^{2}$) & 0.0109(9)    & 0.0102(9)      & 0.0077(7)  & 0.0138(7) ~     & 0.0083(7)  & 0.0121(7)~   & 0.0110(9)    & 0.0191(9) \\
  ~~~~ U$_{33}$ (\AA$^{2}$)          & 0.0462(9)    & 0.0115(9) ~    & 0.1501(9)  & 0.02301(9)      & 0.1628(9)  & 0.0265(9)~   & 0.1765(9)    & 0.0311(9) \\
  NP\tablenote{NP refers to nanoparticle.} diameter (nm) & $\infty$  & $\infty$ &  3.7(1)& 3.7(1)   &  3.1(1)    &  3.1(1)      & 2.4(2)       & 2.2(2) \\
 R$_{w}$~                            & 0.12         & 0.09           & 0.20       & 0.14            &   0.18     &  0.15        & 0.27         & 0.21     \\
\hline
\end{tabular}
\label{tab;rwPDFwur}
\end{table*}

 Attempts to characterize the structure changes using direct
measurements such as TEM technique for such small CdSe
nanoparticles~\cite{lande;nanol01} were unsuccessful due to the poor
contrast.  However, in the present study we were successful in
exploring the local atomic structure for CdSe nanoparticles, in real
space, at different length scales. The PDF fits clearly indicate
that the structure can be described in terms of locally distorted
wurtzite structure containing $\sim 50$\% stacking fault density
(i.e., intermediate between wurtzite and zinc-blende) even for the
2~nm diameter particles, Fig.~\ref{fig:coreFITsfd}.

Interestingly, there is little evidence in our data for a significant
surface modified region.  This surface region is sometimes thought of
as being an amorphous-like region.  Amorphous structures appear in the
PDF with sharp first neighbor peaks but rapidly diminishing and
broadening higher neighbor peaks.  Thus, in the presence of a surface
amorphous region, we might expect to see extra intensity at the
first-peak position when the wurtzite model is scaled to fit the
higher-$r$ features coming just from the crystalline core.  As evident
in Fig.~\ref{fig:coreFITsfd}, this is not observed. Furthermore, as we
describe below, the diameter of the crystalline core that we refine
from the PDF agrees well with other estimates of nanoparticle size,
suggesting that there is no surface amorphous region in these
nanoparticles.  The good agreement in the intensity of the first PDF
peak also presents a puzzle in the opposite direction since we might
expect surface atoms to be under-coordinated, which would result in a
decrease in the intensity of this peak.  It is possible that the
competing effects of surface amorphous behavior and surface under
coordination perfectly balance each other out, and this cannot be
ruled out, though it seems unlikely that it would work perfectly at
all nanoparticle diameters.  This is also not supported by the
nanoparticle size determinations described below.


\subsection{Nanoparticle size}
We describe here how we extracted more accurate
 nanoparticle diameters.  This determination is important since the physical
 proprieties are size dependent.  It is also important
 to use complementary techniques to determine particle size as
 different techniques are more dependent on different aspects of the
 nanoparticle structure, for example, whether or not the technique is
 sensitive to any amorphous surface layer on the nanoparticle. More
 challenges are expected in accurate size determination as
 nanoparticle diameter decreases, due to poor contrast near the
 surface of the nanoparticle.

In the literature, CdSe nanoparticles with a diameter of 2.0~nm have
been considered to be an especially stable size with an associated
band edge absorption centered at 414~nm~\cite{chen;jacs05}, that
size was observed earlier~\cite{murra;jacs93,qu;nanol01} with an
estimated diameter of $\leq$2.0~nm.  There are some reported
difficulties in determining the diameter of such small CdSe
nanoparticles. Attempts to characterize the structure changes by TEM
and X-ray diffraction techniques~\cite{lande;nanol01} were
unsuccessful due to the small diameter of the particles relative to
the capping material.

If we assume the nanoparticle to have spherical shape (a reasonable
approximation based on the TEM in Fig.~\ref{fig:cdse1200tem}) cut
from the bulk, then the measured PDF will look like the PDF of the
bulk material that has been attenuated by an envelope function given
by the PDF of a homogeneous sphere, as follows~\cite{kodam;aca06}
\begin{equation}
\centering
\label{equ;eqGrofSpher}
G\left(r,d\right)_{s} = G\left(r\right)f\left(r,d\right),
\end{equation}
where $G(r)$ is given in Eq.~\ref{equ;eqGrf}, and $f(r,d)$ is a sphere
 envelope function given by
\begin{equation}
\centering
\label{equ;eqSpherformfactor}
f\left(r,d\right) = \left[1-\frac{3r}{2d}+
\frac{1}{2}\left(\frac{r}{d}\right)^3\right]\Theta(d-r),
\end{equation}
where $d$ is the diameter of the homogeneous sphere, and $\Theta(x)$
is the Heaviside step function, which is equal to 0 for negative $x$ and 1
for positive.

\begin{figure}[tbp]
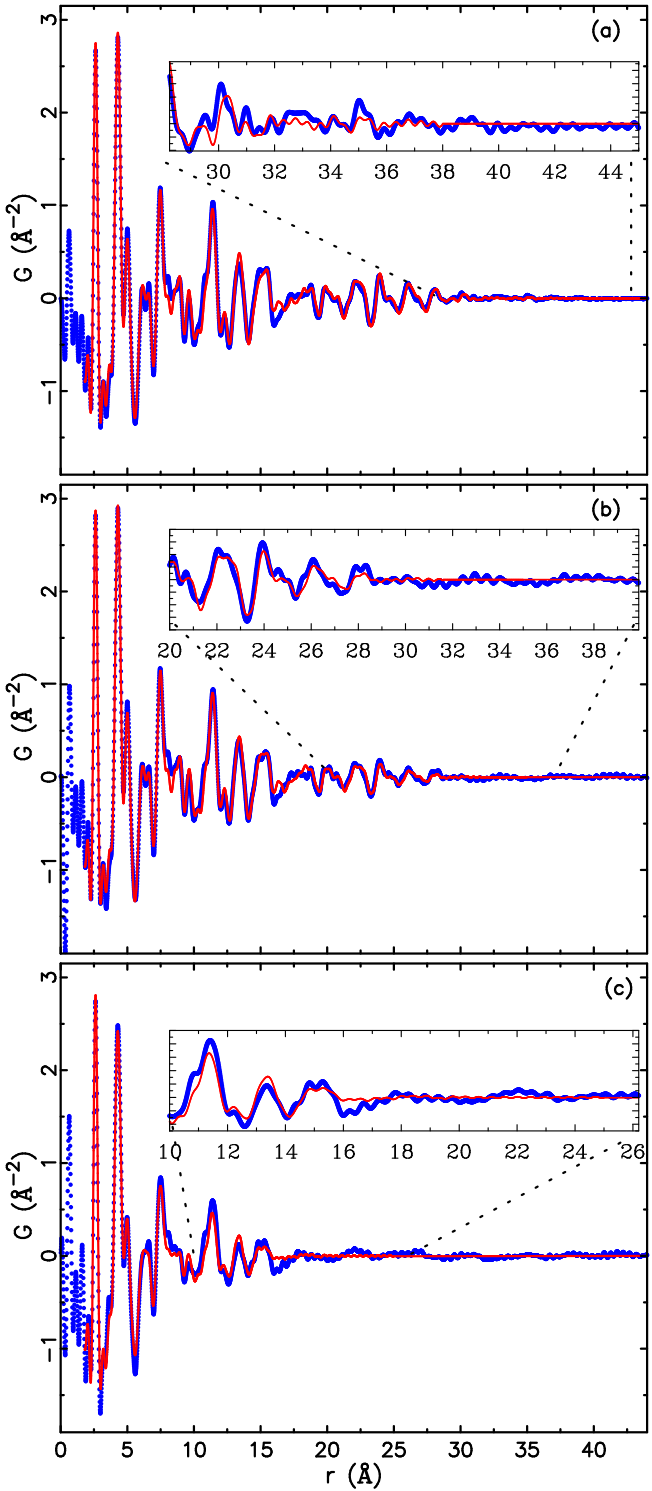

\includegraphics[width=0.48\textwidth]{fig9_a.ps}
\includegraphics[width=0.48\textwidth]{fig9_b.ps}
\includegraphics[width=0.48\textwidth]{fig9_c.ps}
\caption {(Color online) The experimental PDF, $G(r)$, shown as solid
dots. Sphere envelope function (Eq.~\ref{equ;eqSpherformfactor}) is
used to transform the calculated PDF of bulk CdSe, using wurtzite
structure containing 50\% stacking fault density, to give a best fit replication of the
PDF of CdSe nanoparticles (red solid line).  The inset shows on an
expanded scale for the high-$r$ region of experimental $G(r)$ on the
top of simulated PDF data for different diameters of CdSe nanoparticles
(solid line). (a) CdSeIII, (b) CdSeII, (c) CdSeI.  Dashed lines are
guides for the eye. }
\label{fig:NPZ}
\end{figure}

The approach is as follows.  First we refine the bulk CdSe data
using PDFfit.  This gives us a measure of the PDF intensity
fall-off due to the finite resolution of the
measurement.~\cite{egami;b;utbp03} Then the measured value of the
finite resolution was kept as an unrefined parameter after that,
while all the other structural and non structural parameters were
refined. To measure the PDF intensity fall-off due to the finite
particle size, the refined PDF is attenuated, during the
refinement, by the envelope function
(Eq.~\ref{equ;eqSpherformfactor}) which has one refined parameter,
the particle diameter. The fit results are shown in
Fig.~\ref{fig:NPZ} and the resulting values of particle diameter
from the PDF refinement are recorded in
Table~\ref{tab;NPZtemPDF}.
The insets show the calculated and measured PDFs on an expanded
scale. The accuracy of determining the nanoparticle size can be
evaluated directly from this figure.  Features in the measured PDFs
that correspond to the wurtzite structure are clearly seen
disappearing smoothly attenuated by the spherical PDF envelope
function. The procedure is least successful in the smallest
nanoparticles, where the spherical particle approximation on the
model results in features that extend beyond those in the data.  In
this case, the spherical approximation may not be working so well.

The particle diameters determined from the PDF are consistent with
those obtained from TEM, UV-vis and photoluminescence measurements.
In particular, an accurate determination of the average diameter of
the smallest particles is possible in the region where UV-vis and
photoluminescence measurements lose their
sensitivity.~\cite{nanda;cm00} In this analysis we have not considered
particle size distributions, which are small in these materials.  The
good agreement between the data and the fits justify this, though some
of the differences at high-$r$ may result from this and could
contribute an error to the particle size.  Several additional fits to
the data were performed to test the sphericity of the
nanoparticles. Attempts were made to fit the PDF with oblate and
prolate spheroid nanoparticle form factors. These fits resulted in
ellipticities very close to one, and large uncertainties in the
refined ellipticity and particle diameters, which suggests that the
fits are over-parameterized. Another series of fits attempted to
profile the PDF with a lognormal distribution of spherical
nanoparticles. Allowing the mean nanoparticle diameter and lognormal
width to vary resulted in nonconvergent fits, which implies that the
particle sizes are not lognormal distributed.  Therefore, there
appears to be little evidence for significant ellipticity, nor a
significant particle size distribution, as fits assuming undistributed
spherical particles give the best results.

The simple fitting of a wurtzite structure with $\sim 50$\% SFD to the
data will result in an estimate of the coherent structural core of the
nanoparticle that has a structure can be described by a mixture of crystalline structures
intermediate between zinc-blend and wurtzite. Comparing the
nanoparticle core diameter extracted from PDF analysis with the
diameter determined from the standard characterization methods yields
information about the existence of a surface amorphous region. The
agreement between the core diameter extracted from PDF and that
determined from the standard methods (Table~\ref{tab;NPZtemPDF}),
indicates that within our measurement uncertainties, there is no
significant heavily disordered surface region in these nanoparticles,
even at the smallest diameter of 2~nm (Fig.~\ref{fig:NPZ}). In
contrast with ZnS nanoparticles~\cite{gilbe;s04} where the heavily
disordered surface region is about 40\% of the nanoparticle diameter
for a diameter of 3.4~nm, the surface region thickness being around
1.4~nm.~\cite{gilbe;s04}

\subsection{Internal strain }
\begin{figure}[tbp]
\includegraphics[angle=-90, width=0.48\textwidth]{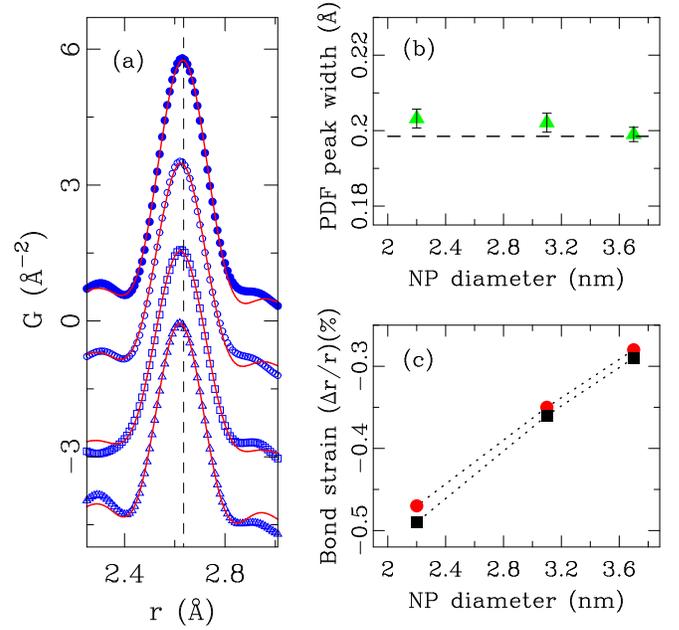}
\caption { (a) The first PDF peak, (\textcolor{blue}{$\bullet$}) bulk,
   (\textcolor{blue}{$\circ$}) CdSeIII, (\textcolor{blue}{$\square$})
   CdSeII and (\textcolor{blue}{$\triangle$}) CdSeI fitted with one
   Gaussian (\textcolor{red}{---}). Dashed line represents the
   position of first PDF peak in the bulk
   data. (b)(\textcolor{green}{$\blacktriangle$}) The first PDF peak
   width vs nanoparticle size, obtained from one Gaussian fit. Dashed line
   represents the width of first PDF peak in the bulk data. (c) Strain
   in Cd-Se bond ($\Delta$$r$/$r$)(\%) vs nanoparticle size.
   (\textcolor{black}{$\blacksquare$}) Bond values obtained from the
   local structure fitting and ({\textcolor{red}{$\bullet$}}) obtained
   from one Gaussian fit to the first PDF peak. Dotted curves are
   guides for the eye. }

\label{fig:strain}
\end{figure}

The local bonding of the tetrahedral Cd-Se building unit was
investigated vs nanoparticle diameter. The nearest neighbor peaks at
$r=2.6353(3)$~\AA\ come from covalently bonded Cd-Se pairs. The
positions and the width of these peaks have been determined by fitting
a Gaussian (Fig.~\ref{fig:strain}(a))
and the results presented in Table~\ref{tab;firstPeakAnalysis}. The
results indicate that there is a significant compressive strain on
this near-neighbor bond length, and it is possible to measure it
with the PDF with high accuracy. The bond length of Cd-Se pairs
shorten as nanoparticle diameter decreases, suggesting the presence
of an internal stress in the nanoparticles. The Cd-Se bond lengths
extracted from the PDF structural refinement are also in good
agreement with those obtained from the first peak Gaussian fit, as
shown in Fig.~\ref{fig:strain}(c).  Thus we have a model independent
and a model dependent estimate of the strain that are in
quantitative agreement.
%
%
\begin{table}
\caption{ The first PDF peak position (FPP) and width (FPW) for
different CdSe nanoparticle sizes and the bulk. }
\begin{tabular}{lcccccccc}  
\hline
 \hline
     &CdSe-bulk & CdSeIII &CdSeII & CdSeI   \\
 \hline
  PDF FPP (\AA) & 2.6353(3)  & 2.6281(3)   &  2.6262(3)  &  2.6233(3)  \\
 \hline
 PDF FPW (\AA)  &  0.1985(09) &  0.1990(19)  &   0.2021(25) &  0.2032(25)  \\
\hline

 \hline
\end{tabular}
\label{tab;firstPeakAnalysis}
\end{table}
The widths of the first PDF peaks have also been extracted vs
nanoparticle diameter from the Gaussian fits
(Table~\ref{tab;firstPeakAnalysis}).  They remain comparably sharp as
the nanoparticles get smaller, as shown in
Fig.~\ref{fig:strain}(b). Apparently there is no size-dependent
inhomogeneous strain measurable on the first peak.  However, peaks at
higher-$r$ do indicate significant broadening
(Fig.~\ref{fig:allrawfqgr}(b)) suggesting that there is some
relaxation taking place through bond-bending.  This is reflected in
enlarged thermal factors
that are refined in the nanoparticle
samples. This is similar to what is observed in semiconductor alloys
where most of the structural relaxation takes place in relatively
lower energy bond-bending distortions.~\cite{petko;prl99,jeong;prb01}

\section{conclusion}
The PDF is used to address the size and structural characterization of
a series of CdSe nanoparticles prepared by the method mentioned in the
text. The core structure of the measured CdSe nanoparticles was found
to possess a well-defined atomic arrangement that can be described in
terms of locally disordered wurtzite structure that contains $\sim
50$\% stacking fault densit, and quantitative structural parameters
are presented.

The diameter of the CdSe nanoparticles was extracted from the PDF data
and is in good agreement with the diameter obtained from standard
characterization methods, indicating that within our measurement
uncertainties, there is no significant heavily disordered surface
region in these nanoparticles, even at the smallest diameter of 2~nm
. In contrast with ZnS nanoparticles~\cite{gilbe;s04} where the heavily disordered
surface region is about 40\% of the nanoparticle diameter for a
diameter of 3.4~nm, the surface region thickness being around
1.4~nm.~\cite{gilbe;s04}

Compared with the bulk PDF, the nanoparticle PDF peaks are broader in
the high-$r$ region due to strain and structural defects in the
nanoparticles.
The nearest neighbor peaks at $r=2.6353(3)$~\AA\, which come from
covalently bonded Cd-Se pairs, shorten as nanoparticle diameter
decreases resulting in a size-dependent strain on the Cd-Se bond
that reaches 0.5\% at the smallest particle size.

\begin{acknowledgments}
We would like to acknowledge help from Didier Wermeille, Doug
Robinson, Mouath Shatnawi, Moneeb Shatnawi and He Lin for help in
collecting data. We are grateful to Christos Malliakas for the
valuable assistance with the transmission electron microscopy. May
thanks to HyunJeong Kim for useful discussion. We are grateful to
Prof. Reinhard Neder for the valuable help with the stacking fault
simulation.  This work was supported in part by National Science
Foundation (NSF) grant DMR-0304391.  Data were collected at the 6IDD
beamline of the MUCAT sector at the Advanced Photon Source (APS). Use
of the APS is supported by the U.S. DOE, Office of Science, Office of
Basic Energy Sciences, under Contract No. W-31-109-Eng-38. The MUCAT
sector at the APS is supported by the U.S. DOE, Office of Science,
Office of Basic Energy Sciences, through the Ames Laboratory under
Contract No. W-7405-Eng-82.

\end{acknowledgments}



\end{document}